\documentclass[twoside]{dis04}
\def\runauthor{N.N. Nikolaev, W. Sch\"afer, B.G. Zakharov, V.R. Zoller}
\def\shorttitle{Color dipoles and $k_\perp$--factorization for nuclei}
\begin{document}

\title{COLOR DIPOLES AND $\bf{k}_\perp$--FACTORIZATION
FOR NUCLEI}

\author{N.N. Nikolaev,{\underline{W. SCH\"AFER}}$^1$,  B.G. Zakharov, and V.R. Zoller}

\address{$^1$ Institut f\"ur Kernphysik \\
Forschungszentrum J\"ulich, D-52425 J\"ulich\\
E-mail: wo.schaefer@fz-juelich.de }

\maketitle

\abstracts{We discuss applications of the color dipole approach to
hard processes on nuclei. We focus on the relation to
$k_\perp$--factorisation and the role of a nuclear unintegrated
gluon distribution in single-- and two-- particle inclusive
spectra in $\gamma^*$A and pA collisions. Linear $k_\perp$
factorisation is broken for a wide class of observables, which we
exemplify on the case of heavy quark $p_\perp$--spectra. }

\noindent {\bf{1. Color dipoles, the unintegrated gluon
distribution and DIS}}\\
When studying the interactions of a highly energetic (virtual)
photon it is of great help to think of its hadronic vacuum
fluctuations as being components of its (light--cone--) wave
function \cite{Gribov-BKS}. Deep inelastic scattering (DIS) can
then be viewed as an interaction of frozen multi--parton Fock
states of the virtual photon with the target nucleon or nucleus.
The proper formalisation valid for inclusive, as well as
diffractive deep inelastic processes results in the color--dipole
approach to small--$x$--DIS \cite{color-dipoles}. Specifically,
the total virtual photoabsorption cross section takes the
well-known quantum mechanical form
$\sigma_{tot}(\gamma^*p;x,Q^2) = \int_0^1 dz \int \,\,\,\,
d^2{\bf{r}} \Psi^*_{\gamma^*}(z,{\bf{r}}) \, \sigma_2 ({\bf{r}})
\Psi_{\gamma^*}(z,{\bf{r}}) \, ,$
with $x,Q^2$ the standard DIS--variables, $\Psi_{\gamma^*}$ is the
$q\bar{q}$--lightcone-wavefunction of the virtual photon, $z,1-z$
are the photon's lightcone momentum fractions carried by the
quark/antiquark, and finally $\sigma_2({\bf{r}})$ is the
dipole--nucleon cross section. The connection between
color--dipole formulas and ${\bf{k}}_\perp$--factorization is
provided by
$\sigma_2({\bf{r}}) =  \sigma_0 \, \int
d^2{\mbox{\boldmath$\kappa$}} [1 -
e^{i{\mbox{\boldmath$\kappa$}}{\bf{r}}}]
f({\mbox{\boldmath$\kappa$}})$, where
$f({\mbox{\boldmath$\kappa$}})$ is directly related to the
unintegrated gluon distribution $ f({\mbox{\boldmath$\kappa$}}) =
{4 \pi \alpha_S \over N_c \sigma_0}\,  {1\over
{\mbox{\boldmath$\kappa$}}^4} \, \partial
G_N/\partial\log({\mbox{\boldmath$\kappa$}}^2)$. Now, for DIS off
nuclei, the dipole is coherent over the whole nucleus for $x \leq
x_A =1/m_N R_A$, where $m_N$ is the nucleon mass, and $R_A$ the
nuclear radius. The dipole--nucleus cross section assumes the
Glauber--Gribov form \cite{color-dipoles} $\sigma_A({\bf{r}}) = 2
\int d^2{\bf{b}} \Gamma_A[\sigma_2({\bf{r}});{\bf{b}}]$, with
$\Gamma_A[\sigma_2({\bf{r}});{\bf{b}}]= 1 -\exp[-{1\over 2}
\sigma_2({\bf{r}}) T_A({\bf{b}})]$, $T_A({\bf{b}})$ is the nuclear
thickness. If we now write $\Gamma_A[\sigma_2({\bf{r}});{\bf{b}}]=
\int d^2{\mbox{\boldmath$\kappa$}}
[1-e^{i{\mbox{\boldmath$\kappa$}}{\bf{r}}}]
\phi({\mbox{\boldmath$\kappa$}})$, then the function
$\phi({\mbox{\boldmath$\kappa$}})$(we suppress its dependence on
${\bf{b}}$) walks and talks like an unintegrated gluon
distribution in inclusive as well as diffractive DIS on nuclei
\cite{NSS}, hence its name 'nuclear unintegrated glue'. It
includes multiple scatterings and the features of nuclear
absorption, as well as ${\bf{k}}_\perp$ broadening of propagating
partons, both controlled by the saturation scale $Q^2_A$. Its
salient features, including a Cronin enhancement at intermediate
${\mbox{\boldmath$\kappa$}}$ and an explicit representation in
terms of convolutions of its free nucleon counterpart can be found
in \cite{NSS,Nonlinear}. Below we shall have a look at the role of
the nuclear unintegrated glue in a broader class of hard,
pQCD--observables than just DIS.

\noindent{\bf{2.Single-- and two particle--inclusive spectra,
$\bf{p}_\perp$--dependence of heavy quarks, and the breakdown of
linear $\bf{k}_\perp$--factorisation}}\\
We now present the essentials of the color--dipole formalism that
allow us to calculate single-- and two--particle spectra
differential in the relevant transverse momenta, as well as e.g.
associated azimuthal asymmetries. Here we think of a situation,
where a highly energetic virtual particle (parton) $a$ dissociates
into two partons, $a\to bc$, in a collision with a heavy nucleus.
The $abc$--coupling should be weak, so that to the first order in
a perturbative coupling (which we absorb into the light--cone wave
function $\Psi({\bf{r}})$ for the $a\to bc$ transition), the
free--particle state is $|a\rangle_{phys} = |a\rangle_0 +
\Psi({\bf{r}}) |bc\rangle_0$, with ${\bf{r}}$ the transverse
distance between $b$ and $c$. The virtue of the impact parameter
representation is the simplicity of the $S$--matrix action on the
bare partons, namely we can write for the scattered wave
\begin{eqnarray}
\textsf{S}|a\rangle_{phys} &=& S_a({\bf{b}}) |a\rangle_0 +
S_b({\bf{b}}_+)S_{c}({\bf{b}}_-) \Psi({\bf{r}}) |bc\rangle_0
\nonumber \\
&=& S_a({\bf{b}}) |a\rangle_{phys} +
\Big[S_b({\bf{b}}_+)S_c({\bf{b}}_-) -S_a({\bf{b}})\Big]
\Psi({\bf{r}}) |bc\rangle_{phys}
\end{eqnarray}
The meaning of the transverse coordinates ${\bf{b}},{\bf{b}}_\pm$
is obvious from Fig 1. Here the terms in brackets represent the
amplitude for the inelastic excitation $a \to bc$, and we may
further identify $S_aS_b$ as a contribution from a scattering of
the constituents, after the dissociation, and $S_a$ as a
contribution of scattering of the parton $a$ before the
dissociation vertex.
Upon squaring the amplitude and
using closure on the nuclear side, one obtains the following form
of the differential, two--particle inclusive cross section for the
process $a\to b({\bf{p}}_+) c({\bf{p}}_-)$:
\begin{eqnarray}
{(2 \pi)^4 d\sigma \over dz d^2{\bf{p}}_+ d^2{\bf{p}}_-} = \int
d^8\Big\{{\bf{b}}_i\Big\} e^{ i{\bf{p}}_+ ({\bf{b}}_+ -
{\bf{b}}_+') + i{\bf{p}}_- ({\bf{b}}_- - {\bf{b}}_-')}
\Psi({\bf{b}}_+ - {\bf{b}}_-)\Psi^*({\bf{b}}'_+ - {\bf{b}}'_-)
\nonumber \\
\Big\{ S^{(4)} ({\bf{b}}_+,{\bf{b}}_-,{\bf{b}}'_+,{\bf{b}}'_-) +
S^{(2)}({\bf{b}},{\bf{b}}') -
S^{(3)}({\bf{b}}_+,{\bf{b}}_-,{\bf{b}}')
-S^{(3)}({\bf{b}}'_+,{\bf{b}}'_-,{\bf{b}})\Big\} \, . \, \, \, \,
\, \, \label{2-PI}
\end{eqnarray}
\begin{figure}[!thb]
\vspace*{3cm}
\begin{center}
\includegraphics{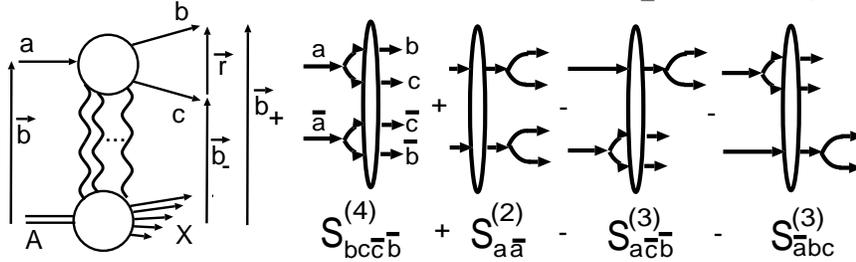}
\caption[*]{{\textsf{Left:}} Amplitude for the process $aA\to
bcX$. Multiple gluon exchanges connect between the nuclear-- and
$a$--fragmentation regions. The relevant impact parameters, which
are conserved in the high energy limit, are indicated.
{\textsf{Right:}} Diagrammatic representation for the evolution
operator of the four parton density--matrix. Particles from the
complex conjugated amplitude become antiparticles in the
four--body  density matrix. Their impact parameters are the primed
ones in the text.}
\end{center}
\vspace{-0.5cm}
\end{figure}
Here the integration is over the impact parameters ${\bf{b}}_\pm,
{\bf{b}}'_\pm$, ${\bf{b}}= z {\bf{b}}_+ + (1-z) {\bf{b}}_-$, and
${\bf{b}}'= z {\bf{b}}'_+ + (1-z) {\bf{b}}'_-$ where $z$ is the
fraction of $a$'s light cone momentum carried by $b$. Here
$S^{(4,3,2)}$ is an appropriate matrix element of the intranuclear
evolution operator for a four(three,two)--particle system, coupled
to an overall color--singlet state, cf.{\cite{Nonlinear}}. We
stress that the intranuclear evolution operator is a matrix in the
space of singlet four-parton dipole states $|R{\bar{R}}\rangle
=|(bc)_R \otimes (\bar{b}\bar{c})_{\bar{R}}\rangle$, further
details depend on the color representations of the partons
involved, e.g. $R=1,8$ for $bc=q\bar{q}$,
$R=1,8_A,8_S,10+{\bar{10}},27$ for $bc=gg$. A further evaluation
of $S^{(4,3,2)}$ would involve the standard Glauber--Gribov
approximation for a dilute gas nucleus of color--singlet nucleons.
For the scattering off individual nucleons the two--gluon exchange
approximation is certainly appropriate in a range of typical
Bjorken $x$ not much lower than $x_A$ (i.e. the range $10^{-3}\leq
x \leq 10^{-2}$ relevant for RHIC or a possible future
electron--nucleus collider \cite{EIC}). It is important to realize
that the color coupled channel aspect of the intranuclear dipole
evolution {\emph{cannot}} be absorbed into a single,
'color--scalar' unintegrated gluon distributions of the nucleus.
Hence, for two--particle--inclusive spectra {\emph{there is no
${\bf{k}}_\perp$--factorization}. Instead, depending again on the
color multiplets of the multiparton system that interacts
coherently with the nucleus, multigluon exchange effects call upon
a whole density matrix of nuclear gluons in color space.
Single--particle inclusive spectra (for a host of examples see e.g
\cite{NSS,single-particle}) are in most relevant cases of abelian
nature and transitions between color channels during intranuclear
rescattering do not appear. Still, if the dissociating particle
$a$ interacts with the nucleus by gluon exchanges,
${\bf{k}}_\perp$--factorization is violated already in the
single--particle spectra. We make our point on the example of the
transverse--momentum spectrum of heavy quarks in $pp$ and
$pA$--collisions, thereby generalizing \cite{NPZ}. For the free
nucleon target eq.(\ref{2-PI}) reduces to (${\bf{p}}$ is the
transverse momentum of the heavy quark $Q$):
\begin{eqnarray}
{2(2\pi)^2 d\sigma(g^* N\to Q\bar{Q} X) \over dz d^2{\bf{p}}}
 =
\int d^2{\bf{r}} d^2{\bf{r}}' e^{i{\bf{p}}({\bf{r}}-{\bf{r}}')}
\Psi({\bf{r}}) \Psi^*({\bf{r}}')
\Big\{\sigma_3(z{\bf{r}}',{\bf{r}})
+ \sigma_3(z{\bf{r}},{\bf{r}}') \nonumber \\
- \sigma_{2,Q\bar{Q}} ({\bf{r}}-{\bf{r}}') -
\sigma_{2,gg}(z({\bf{r}}-{\bf{r}}'))\Big\} \, , \,\,\,\,\,\,\,
\end{eqnarray}
with the three--body dipole cross section $
\sigma_3({\bf{x}},{\bf{r}}) = {C_A\over 2C_F}
[\sigma_2({\bf{x}})+\sigma_2({\bf{x}}-{\bf{r}}) - {1 \over N_c^2}
\sigma_2({\bf{r}}) ] \equiv  {\cal{F}}[\sigma_2]$, and
$\sigma_{2,gg}({\bf{x}}) = {C_A \over C_F}
\sigma_{2,Q\bar{Q}}({\bf{x}})$. We indicated that for the free
nucleon target $\sigma_3$ is a certain linear functional
${\cal{F}}$ of the two--body dipole cross section, and thus also
of the unintegrated gluon distribution. Now, when going to the
nuclear target, we utilize the Glauber--Gribov substitution
$\sigma_2({\bf{r}}) \to \sigma_{2A} ({\bf{r}}) = 2 \int
d^2{\bf{b}} \Gamma_A[\sigma_2({\bf{r}});{\bf{b}}];\,
\,\sigma_3({\bf{x}},{\bf{r}}) \to \sigma_{3A} ({\bf{x}},{\bf{r}})
= 2 \int d^2{\bf{b}}
\Gamma_A[\sigma_3({\bf{x}},{\bf{r}});{\bf{b}}]$
and, obviously, $\sigma_{3A}$ is not the same linear functional of
$\sigma_{2A}$ as its free--nucleon counterpart: $\sigma_{3A} \neq
 {\cal{F}}[\sigma_{2A}]$. Thus, the single--particle inclusive
transverse momentum spectrum of heavy quarks in $pA$--collisions
is necessarily a different functional of the nuclear unintegrated
glue than the corresponding spectrum in $pp$ collision is of the
proton's unintegrated glue. In short: ${\bf{k}}_\perp$
factorization does not hold. This seemingly somewhat technical
point is maybe best illustrated by a look at momentum space
formulas. The free--nucleon cross section now becomes
\begin{eqnarray}
{2(2\pi)^2 d\sigma(g^* N\to Q\bar{Q} X) \over dz d^2{\bf{p}}}
 = \int d^2{\mbox{\boldmath$\kappa$}} f({\mbox{\boldmath$\kappa$}})\Big\{{C_A\over 2 C_F} \Big( |\Psi({\bf{p}} )
 -\Psi({\bf{p}}+z {\mbox{\boldmath$\kappa$}})|^2 \nonumber \\
 + |\Psi({\bf{p}}+{\mbox{\boldmath$\kappa$}}) - \Psi({\bf{p}}+z{\mbox{\boldmath$\kappa$}})|^2 - |\Psi({\bf{p}})-\Psi({\bf{p}}+{\mbox{\boldmath$\kappa$}})|^2
  \Big) + |\Psi({\bf{p}})-\Psi({\bf{p}}+{\mbox{\boldmath$\kappa$}})|^2 \Big\} \, ,
\label{linear}
\end{eqnarray}
where the linear dependence on the unintegrated glue
$f({\mbox{\boldmath$\kappa$}})$ is in clear evidence. If
eq.(\ref{linear}) was a true factorization theorem, all the target
dependence would be buried in $f$, and the nuclear cross section
should just be given by properly substituting $f \to \phi$.
Instead, in a strong absorption regime, say for central
$g^*$--nucleus collisions, the nuclear cross section has a
drastically different functional dependence on the (nuclear--)
unintegrated glue, namely:
\begin{equation}
{(2\pi)^2 d\sigma(g^* A\to Q\bar{Q} X) \over dz d^2{\bf{p}}
d^2{\bf{b}}}\Big|_{{\bf{b}}\to 0} = \int
d^2{\mbox{\boldmath$\kappa$}}_1 d^2{\mbox{\boldmath$\kappa$}}_2
\phi({\mbox{\boldmath$\kappa$}}_1)
\phi({\mbox{\boldmath$\kappa$}}_2)
|\Psi({\bf{p}}+{\mbox{\boldmath$\kappa$}}_2)
-\Psi({\bf{p}}+z{\mbox{\boldmath$\kappa$}}_1+z{\mbox{\boldmath$\kappa$}}_2)
|^2 \label{nonlinear}
\end{equation}
It is important to stress that the nonlinear (quadratic)
dependence of the heavy quark spectrum on the target's
unintegrated glue has nothing to do with matters of taste
concerning our definition of a gluon distribution. Simply, with
$f$ and $\phi$ both defined by means of the same observable --the
total DIS--cross section--equations (\ref{linear},\ref{nonlinear})
entail a very different {\emph{relation between two
observables}}--the total DIS cross section and the heavy quark
spectrum--depending on whether the target is a single nucleon or a
strongly absorbing nucleus. We conclude that phenomenologies which
treat hard nuclear processes simply by substituting nuclear gluon
distributions into linear $\bf{k}_\perp$--factorization formulas
are not borne out by a consistent treatment, and certainly have
nothing to say about a possible role of
saturation/absorption/multiple scattering effects in
hadron--nucleus collisions. Finally, we remark that in a limit
where ${\bf{p}}$ becomes the hardest scale, our
eq.(\ref{nonlinear}) smoothly connects to the much cherished hard
collinear factorisation theorems. A similar phenomenon is observed
for gluon jets in $g^* \to gg$, where a cubic dependence on $\phi$
is obtained in the strong absorption limit. Violations of linear
${\bf{k}}_\perp$--factorisation had previously been discussed in
the breakup of virtual photons into dijets \cite{Nonlinear}
$\gamma^* \to q\bar{q}$, where the correct treatment of the color
multichannel aspects is crucial. There we found a complete
azimuthal decorrelation of semihard dijets with transverse momenta
below the saturation scale, in which case the relation to the
nuclear unintegrated glue is highly nonlinear. For hard dijets a
linear dependence on the unintegrated glue emerges, with
${\bf{k}}_\perp$--factorisation however being violated. Still,
sizeable jet decorrelation effects from intranuclear rescattering
remain, especially in central DIS. 
\end{document}